\documentstyle[preprint,aps]{revtex}
\begin{document}
\draft
\preprint{NCL93-TP12}

\title{No generalized statistics from dynamics in curved spacetime}

\author{J. W. Goodison and D. J. Toms}
\address{Physics Department, University of Newcastle upon Tyne,\\ Newcastle
upon Tyne, NE1 7RU, U.K.}
\date{November 10, 1993}

\maketitle

\begin{abstract}
  We consider quon statistics in a dynamically evolving curved spacetime in
which prior to some initial
  time and subsequent to some later time is flat. By considering the Bogoliubov
transformations associated with
  gravitationally induced particle creation, we find that the consistent
evolution of the generalized
  commutation relations from the first flat region to the second flat region
can only occur if $q = \pm 1$.

\vskip 0.5in

PACS numbers: 05.30.-d, 04.60.+n

\vskip 1in
\center{ ( To appear in Physics Review Letters ) }

\end{abstract}

\pacs{}

\narrowtext

\section*{ }

  There has been an increasing interest in generalized commutation relations of
the form

\begin{equation}
a_{i} a_{j}^{\dagger} - q a_{j}^{\dagger} a_{i} = \delta_{ij}
\end{equation}
  where $q$ is real, coming from two main directions. One motivation comes from
the study of quantum groups
  ( for example see Refs. \cite{Jim1,Jim2,Jim3,Wor2,Wor3} );
  in particular the quantum version \cite{Bied,Macf} of SU(2). The other
motivation is Greenberg's study of particles
  with infinite statistics \cite{GreenPRL,GreenPRD}. In either case, the
relations (1) can be used to quantize
  the simple
  harmonic oscillator. ( The generalized commutation relations used in Refs.
\cite{Bied,Macf}
   can be put in the form (1) by
  an appropriate transformation. ) As $q$ ranges from +1 to -1, it can be seen
that the relations (1)
  interpolate between commutation relations and anticommutation relations
  ( appropriate for bosons and fermions respectively ).

  Given an operator $a_i$ and its adjoint $a_i^{\dagger}$ which satisfy (1), a
vacuum state $| 0 >$
  can be defined in the usual way by

\begin{equation}
a_{i} | 0 > = 0
\end{equation}
   and the Fock space built up as normal by applying monomials in
$a_{i}^{\dagger}$ to $| 0 >$. A detailed study of the
  resulting space has been given \cite{Fivel,zag}. It was shown in these
references that for $|q| < 1$, the Fock space
  is well defined with all states of positive norm. An important feature,
emphasised by Greenberg \cite{GreenPRD}
  is that it is not possible to impose any relation which relates $a_{i}a_{j}$
to $a_{j}a_{i}$. A consequence
  of this is that multiparticle states have no definite symmetry properties
under particle interchange.
  In the special case $|q|=1$, it is still true that all vectors in the Fock
space have non-negative
  norms; however, unless the vectors are totally symmetric or antisymmetric
under particle interchange,
  they have zero norm. The allowed cases when $|q|=1$  correspond to the normal
Bose or Fermi statistics.
  If $|q| > 1$ then the Fock space has states of negative norm, meaning that
the Hilbert space structure
  is lost, along with the usual probability interpretation. We will restrict
ourselves to $ |q| < 1 $.

  As far as we are aware, all studies of the relations (1) have been in flat
spacetime. The aim of the present
  paper is to examine the consequences of imposing such relations in curved
spacetime. In particular, we
  will examine Parker's proof of the spin-statistics theorem applied to a
system quantized using (1).
  The basic idea of the proof is to suppose that we have a time dependent
spacetime which for $ t<t_1$
  is flat, and which for $t>t_2$ is also flat. ( Here $t_1$ and $t_2$ are two
arbitrary times. ) The
  spacetime can be dynamic for $ t_1 \le t \le t_2 $. As a specific example we
could consider a spatially
  flat Robertson-Walker spacetime with

\begin{equation}
ds^2 = dt^2 - R^2(t) \bigl( dx^2 + dy^2 + dz^2 \bigr)
\end{equation}
  where $R(t)=R_1$ for $t \le t_1$ and $R(t)=R_2$ for $t \ge t_2$. We will call
the portion of spacetime
  with $t<t_1$ the in-region, and that with $t>t_2$ the out-region. What Parker
originally showed \cite{park1} was
  that for a single real scalar field of spin zero obeying the Klein-Gordon
equation, if commutation relations were
  imposed on the creation and annihilation operators in the in-region, then the
operators had to obey commutation
  relations in the out-region. It is not consistent to impose anticommutation
relations on such a field.
  Parker later generalized the proof to fermions of spin-1/2 which obey the
Dirac equation and satisfy
  anticommutation relations \cite{park2}. The cases of parastatistics
\cite{paw} and ghost fields
  \cite{park4} have also been treated. It was emphasised that
  by taking the limit in which the dynamical part of the evolution becomes
constant, and demanding
  that the commutation relations imposed remain continuous in this limit, the
usual spin-statistics
  relation holds true also in flat spacetime. ( See Sec. B of Ref. \cite{park1}
for a discussion. )

  In the in-region we may expand the field operator $\Phi (x)$ in terms of
creation and annihilation
  operators as

\begin{equation}
\Phi (x) = \sum_i \bigl( F_{i}(x)a_{i} + F_{i}^*(x)a_{i}^{\dagger} \bigr).
\end{equation}
  $\Phi(x)$ will be taken to be a real scalar field in the Heisenberg picture
here, although there
  is nothing inherently relativistic in the argument and an analogous treatment
could be given for the Schr\"odinger
  field. $ \{ F_i (x) \}$ is a complete set of positive frequency solutions to
the Klein-Gordon
  equation. We will assume that the operators in (4) satisfy (1) with $ |q| < 1
$. We note
  that the quantum fields we consider are interacting with a gravitational
field of arbitrary strength, but
  are otherwise free.

  An expansion similar to (4) may be imposed in the out-region:

\begin{equation}
\Phi (x) = \sum_i \bigl( G_{i}(x)b_{i} + G_{i}^*(x)b_{i}^{\dagger} \bigr)
\end{equation}
 with $ \{ G_i(x) \} $ a complete set of positive frequency solutions. The
operators $b_i$ and $b_i^{\dagger}$
  will in general differ from $a_i$ and $a_i^{\dagger}$ if there is particle
creation due to the expansion
  of the universe \cite{park1,park2,park5}. We assume that
\begin{equation}
b_{i} b_{j}^{\dagger} - q^{\prime} b_{j}^{\dagger} b_{i} = \delta_{ij}
\end{equation}
  where $|q^{\prime}| < 1$ with $ q^{\prime}$ not necessarily equal to $q$.

  Because both sets $\{F_{i}(x)\}$ and $\{G_{i}(x)\}$ are assumed to be
complete, we may expand one in terms
   of the other:

  \begin{equation}
  G_{i}(x) = \sum_{j} \bigl( \alpha_{ij}F_j(x) + \beta_{ij}F^*_j(x) \bigr)
  \end{equation}
   for some coefficients $\alpha_{ij}$ and $\beta_{ij}$. The expansion
coefficients in (7) are called
  Bogoliubov coefficients. They were first used to study particle creation in
the expanding universe by
  Parker \cite{park5}. We may now substitute (7) and its complex conjugate into
(5). Comparison of the result
  with (4) shows that

  \begin{equation}
  a_i = \sum_j \bigl( \alpha_{ji}b_j + \beta_{ji}^{*} b_j^{\dagger} \bigr).
  \end{equation}
   We have demanded that $a_i$ and $a_j^{\dagger}$ satisfy (1), whereas $b_i$
and $b_j^{\dagger}$
  satisfy (6). However, for a spacetime whose metric is of the form given in
(3), the Bogoliubov coefficients
  are diagonal. Therefore we restrict ourselves to the case where
\begin{equation}
\alpha_{ij} = \alpha_i \delta_{ij}
\end{equation}
  \begin{equation}
\beta_{ij} = \beta_i \delta_{ij}
\end{equation}
   where there is no sum over the repeated indices.
   Substitution of (8) and its complex conjugate into (1), and using (6) leads
to

\begin{eqnarray}
\delta_{ij} = ( | \alpha_i |^2 - q | \beta_i |^2 ) \delta_{ij} + \alpha_i
\beta_j ( b_i b_j - q b_j && b_i )
+ \beta_i^* \alpha_j^* ( b_i^{\dagger} b_{j}^{\dagger} - q b_j^{\dagger}
b_i^{\dagger} ) \nonumber \\ &&
+ ( 1 - q^{\prime} q ) \beta_i^* \beta_j b_i^{\dagger} b_j + ( q^{\prime} - q )
\alpha_i \alpha_j^* b_j^{\dagger} b_i.
\end{eqnarray}
  This differs considerably from the standard case since it is an operator
relation. In the cases $ q^{\prime}
  = q = \pm 1 $, terms in (11) which involve operators
  automatically vanish on account of the commutation or anticommutation of
  $b_k$ with $b_l$ and $b_k^{\dagger}$ with $b_l^{\dagger}$, corresponding
  to the requirement that the states be symmetric or antisymmetric
  under particle interchange. For $ |q|<1$, the requirement that (11) holds
imposes extra conditions to
  be satisfied which are not found in the usual case.

  Analogously to (2) we define a vacuum state in the out-region by

  \begin{equation}
  b_{i} | 0, {\rm out} > = 0.
  \end{equation}
   ( Similarly $ a_{i} | 0, {\rm in} > = 0 $ defines a vacuum state in the
in-region. )
    Taking the expectation value of (11) with $| 0, {\rm out} >$ leads to

   \begin{equation}
   1 = | \alpha_i |^2 - q | \beta_i |^2.
   \end{equation}
   This is the obvious extension of the results of Parker
\cite{park1,park2,park5} to general $q$ and diagonal
   Bogoliubov transformation.
   Using (13) in (11) gives

\begin{eqnarray}
0 = \alpha_i \beta_j ( b_i b_j - q b_j b_i )
+ \beta_i^* \alpha_j^* ( b_i^{\dagger} b_{j}^{\dagger} - q b_j^{\dagger}
b_i^{\dagger} )
+ ( 1 - q^{\prime} q ) \beta_i^* \beta_j b_i^{\dagger} b_j + ( q^{\prime} - q )
\alpha_i \alpha_j^* b_j^{\dagger} b_i.
\end{eqnarray}
   Because the operators which appear on the RHS of (14) are independent of
each other, the
   coefficients must vanish separately. Another way to see this is to take the
expectation value
   of (14) in states other than the vacuum. If we define

\begin{equation}
| kl, {\rm out} > = b_k^{\dagger} b_l^{\dagger} | 0, {\rm out} >,
\end{equation}
   then it is easy to show, using (6), that

\begin{equation}
< kl, {\rm out} | mn, {\rm out} > = \delta_{km} \delta_{ln} + q^{\prime}
\delta_{lm} \delta_{kn}.
\end{equation}
  It is also easy to show that the states defined in (15) are linearly
independent, since if we have

\begin{equation}
\sum_{k,l} c_{kl} | kl, {\rm out} > = 0
\end{equation}
  for some coefficients $c_{kl}$ then (16) may be used to obtain

\begin{equation}
c_{mn} + q^{\prime} c_{nm} = 0,
\end{equation}
  from which it follows that
\begin{equation}
( 1 - {q^{\prime}}^2 ) c_{mn} = 0.
\end{equation}
  Since we assume $ |q^{\prime}| < 1 $ the coefficients $c_{mn}$ must vanish.

  If we now operate with (14) on $ | 0, {\rm out} > $ and use (12) and (15) we
have

\begin{equation}
0 = \beta_i^* \alpha_j^* ( | ij, {\rm out} > - q | ji, {\rm out} > ).
\end{equation}
  Using the linear independence of the states $| ij, {\rm out} >$, it then
follows that

\begin{equation}
\beta_i^* \alpha^*_j = 0
\end{equation}
  The remaining relation that follows from (14) is

\begin{equation}
0 = ( q^{\prime} - q ) \alpha_{i} \alpha_{j}^* b_j^{\dagger} b_i + ( 1 -
q^{\prime} q ) \beta_{i}^* \beta_{j} b_i^{\dagger} b_j.
\end{equation}
 In particular (22) must hold for $i=j$. Then we obtain

\begin{equation}
0 = \{ ( q^{\prime} - q ) | \alpha_{i} |^2  + ( 1 - q^{\prime} q ) | \beta_{i}
|^2 \} b_i^{\dagger} b_i.
\end{equation}
  Since $b_i^{\dagger} b_i$ cannot vanish as an operator ( e.g. take the
expectation value of (23) with the state
  $ | i, {\rm out } > $ ) then

\begin{equation}
0 = ( q^{\prime} - q ) | \alpha_{i} |^2  + ( 1 - q^{\prime} q ) | \beta_{i} |^2
{}.
\end{equation}
  Now let us suppose that at least one of the $\alpha_i$ vanishes. Then, from
(13), we cannot have $ \beta_i = 0 $.
  Thus we find $q^{\prime} q = 1$ which contradicts $ |q| < 1 $, $ | q^{\prime}
| < 1 $. Therefore none of the $ \alpha_i $
  can vanish.

  This in turn demands, from (21), that all of the $ \beta_i $ must vanish and
since $ \alpha_i \not = 0 $ that $
  q^{\prime} = q $.
  However, the requirement that all $ \beta_{i} = 0$ also leads to a
contradiction. It implies that positive frequency
  solutions in the in-region remain
  positive frequency in the out-region even if the universe undergoes a
dynamical evolution. As in the normal case,
  it corresponds to the absence of particle creation. However, for a spacetime
of the type given in (3), the argument
   given in \cite{park1} holds, and shows that in general $ \beta_{i} \not =
0$.
  ( A number of spacetimes leading to $\beta_{i} \not = 0$ are known; see
\cite{park6,BD} for references. )

  In conclusion we have shown that the assumptions $|q|<1$ and $|q^{\prime}|<1$
lead to an unavoidable contradiction
  in a dynamically evolving universe. There are only two possible alternatives.
  The first is to start with $q=+1$ ( for bosons ) or $q=-1$ ( for fermions )
which is already covered by Parker's
  work \cite{park1,park2}. It is necessary for $q^{\prime}=+1$ ( for bosons )
or $q^{\prime}=-1$ ( for fermions ).
  The second possibility is to start with $q^{\prime}=+1$ ( for bosons ) or
$q^{\prime}=-1$ ( for fermions ). However,
  since the dynamical evolution of the universe in this proof is arbitrary, one
may simply reverse time in Parker's
  work to see that $q=+1$ ( for bosons ) or $q=-1$ ( for fermions ).
   Hence the consistent evolution of the generalized commutation relations
  from the in-region to the out-region may only occur if $q=q^{\prime}=\pm 1$.
We expect that an argument similar to
  that given above may be presented for quons in a background electromagnetic
field which allows the possibility
  of particle creation. In addition to problems pointed out by Greenberg
\cite{GreenPRD}, we expect that our paper
  essentially rules out a quantum field theory based on quons.

\acknowledgements

   JWG would like to thank the S.E.R.C. for financial support. DJT would like
to thank the Nuffield Foundation for their
   support.

\end{document}